# Partially Coherent Radar Unties Range Resolution from Bandwidth Limitations

Rony Komissarov[1,=], Vitali Kozlov[1,=], Dmitry Filonov[1], and Pavel Ginzburg[1,2,*]

[1]School of Electrical Engineering, Tel Aviv University, Tel Aviv, 69978, Israel
[2]Light-Matter Interaction Centre, Tel Aviv University, Tel Aviv, 69978, Israel

**Abstract**:

It is widely believed that range resolution, the ability to distinguish between two closely situated targets, depends inversely on the bandwidth of the transmitted radar signal. Here we demonstrate a different type of ranging system, which possesses superior range resolution that is almost completely free of bandwidth limitations. By sweeping over the coherence length of the transmitted signal, the partially coherent radar experimentally demonstrates an improvement of over an order of magnitude in resolving targets, compared to standard coherent radars with the same bandwidth.. A theoretical framework is developed to show that the resolution could be further improved without a bound, revealing a tradeoff between bandwidth and sweep time. This concept offers solutions to problems which require high range resolution and accuracy but available bandwidth is limited, as is the case for the autonomous car industry, optical imaging, and astronomy to name just few.

**Introduction**

[=] equal contribution
[*] pginzburg@post.tau.ac.il

Spatial and temporal coherence of electromagnetic radiation manifest in the ability of the wave to interfere with itself under interferometric experiments (e.g. observation of interference fringes) [1]. This peculiar statistical property was already utilized almost a century ago by Michelson for performing stellar imaging [2]. Today, optical coherence tomography (OCT) turns the finite coherence length of light sources into an advantage [3-5], performing imaging with deep sub millimeter resolution inside biological tissues. Distances to scattering objects under investigation are deduced by sweeping over the delay between a transmitted and a reflected signal and then recovering the peak of the coherence function. While control over the coherence length of optical sources is still a technological challenge, the low frequency part of the electromagnetic spectrum (below 1THz) has benefited from coherent sources since the invention of the quartz resonator and the homodyne receiver over a century ago [6],[7]. Today, the availability of cheap and reliable electronics allows unprecedented control over the shape of the transmitted electromagnetic fields, as well as precise measurement of the reflected signals' phase. These advantages lead to the dominance of fully coherent sources in most radar implementations as we know them today [8]. Partially coherent sources, on the other hand, have remained largely unexplored below the optical frequency spectrum. A notable exception, termed 'noise radar', employs stochastic degrees of freedom to modulate the carrier wave, e.g. [9-16]. However, the method of operation is different to the presented approach because it does not involve noise..

In order to achieve high range and Doppler resolutions, many types of coherent carrier modulations have been proposed [8],[17], resulting in different radar implementations, such as frequency modulated continuous wave (FMCW), which transmit continuously, and coded pulse radars, which have non transmitting periods. It is important to underline that the word "coherent" in this report does not refer to the integration method, which is commonly used in the context of standard radars, but rather to the property of the transmitted wave itself. At this point it may be asked why bother improving on the well-developed radar systems? While various radar signals offer advantages over others, to the best of the authors' knowledge, all suffer from a link between range resolution (the ability to distinguish between two closely situated targets) and transmitted signal bandwidth, leading to the opinion that it is an unbreakable relation. Furthermore, range accuracy, the certainty with which the distance to a single target is known, tends to have a similar inverse dependence on bandwidth as well [18],[19] (mostly in FMCW realizations). This forces expensive high bandwidth implementations in applications where range accuracy and range resolution are crucial. Typical examples include but are not limited to the automotive and security industries, which can require over 1GHz of bandwidth to achieve resolution of better than a meter [20],[21]. The cost of high bandwidth hardware and the regulation of the allowed spectral bands push towards reducing the beforehand mentioned dependence. A few ideas were proposed to achieve this goal [10],[22-25], but they are mostly limited to special conditions and only offer an improvement by a relatively small factor, leaving the fundamental dependence on bandwidth intact.

Here we propose and experimentally demonstrate a different type of super-resolution ranging and detection system inspired by OCT, which is not limited by bandwidth, at the expense of longer acquisition time. The concept is depicted in Fig. 1, where gradual increase of the coherence length of a CW radiation allows mapping a road scenario, as a representative example.

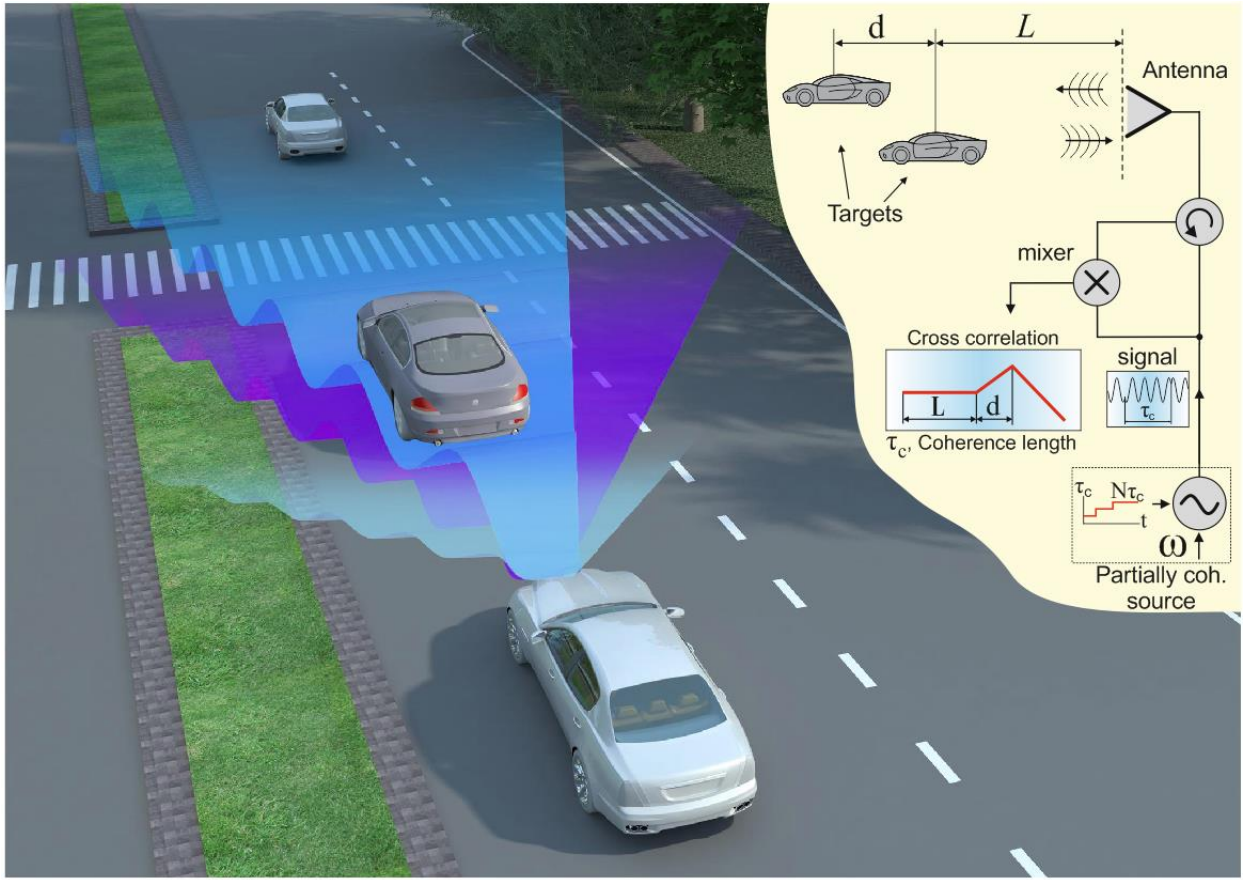

**Figure 1**: Illustration of the partially coherent radar concept. Three different waves are shown: light color – smallest, purple – intermediate and blue - longest coherence lengths. The width of the beam is drawn differently for each wave solely for clarity of illustration. For the lightly colored wave, the reflected signal from the cars is no longer correlated with the still transmitting part of the signal, due to its short coherence length. The purple wave, reflected from the first car, is correlated with the transmitting signal, but the reflections from the second car are not, which allows to detect the distance of the first one. The blue wave has the longest coherence length that correlates with reflections from both objects, allowing the detection of the second car as well. The coherence length (or time) of the radar is swept from shortest to longest, scanning the location of targets along the line of sight. *Inset* - Schematic representation of the radar system. An oscillator with controllable coherence time $\tau_c$ is transmitted and mixed with the reflections from the targets. The phase is switched *N* times and the output of the mixer is averaged over a window of length $N\tau_c$. Repeating the process by sweeping over the coherence length produces the cross correlation as a function of coherence length. The location of the targets is extracted from this data.

# Results

## A. Theory and Implementation

In contrast to optical sources, where coherence is governed by internal noise (typically Gaussian), radio frequency (RF) technology allows control over the coherence almost on demand. This permits an all analogue implementation that can quickly sweep over the coherence length and recover the cross correlation function between the transmitted and received signals. In order to implement this partially coherent source, the phase of the carrier can be switched uniformly in the range of $[0,2\pi]$, where the time between switching events is distributed exponentially with a mean time that is related to the desired coherence length (as in optical sources) [26]. A complementary approach, undertaken here, has technological advantages in RF implementations and suggests generating a 100% duty cycle pulse train (CW), where the random switching time is replaced with constant coherence intervals. This architecture corresponds to deterministic time between random phase jumps, which possess similar cross correlation properties when sweeping over the coherence length. Schematic representation of an implementation of a partially coherent oscillator is shown latter and will be discussed after describing the basic signal processing formalism. The proposed radar signal is given by Eq.1:

$$S(t) = \text{Cos}(\omega t + \varphi(t))$$

$$\varphi(t) = \sum_{n=0}^{N-1} \sum_{m=0}^{M-1} \text{rect}\left(\frac{t - n\tau_m - T_m}{\tau_m}\right) \varphi_{nm}, \quad (1)$$

where $\text{rect}(t) = \begin{cases} 1 & 0 \leq t \leq 1 \\ 0 & o.w. \end{cases}$, $\varphi_{nm}$ are uniformly distributed i.i.d (independent and identically distributed) phases, $\tau_m = \tau_0 + m\frac{\Delta\tau}{M-1}$ is the length of each constant phase pulse (or coherent time duration, which increases after N pulses were sent), $T_m = \begin{cases} N\sum_{q=0}^{m-1} \tau_q & m \geq 1 \\ 0 & o.w. \end{cases}$ is the time that passed from the beginning of the scan until iteration $m$, $(l_0, l_0 + \Delta l) = (c\tau_0, c\tau_0 + c\Delta\tau)$ is the scanned range along the line of sight and $\omega = ck$ is the frequency of the carrier. An illustration of the transmitted signal is shown on Fig.2.

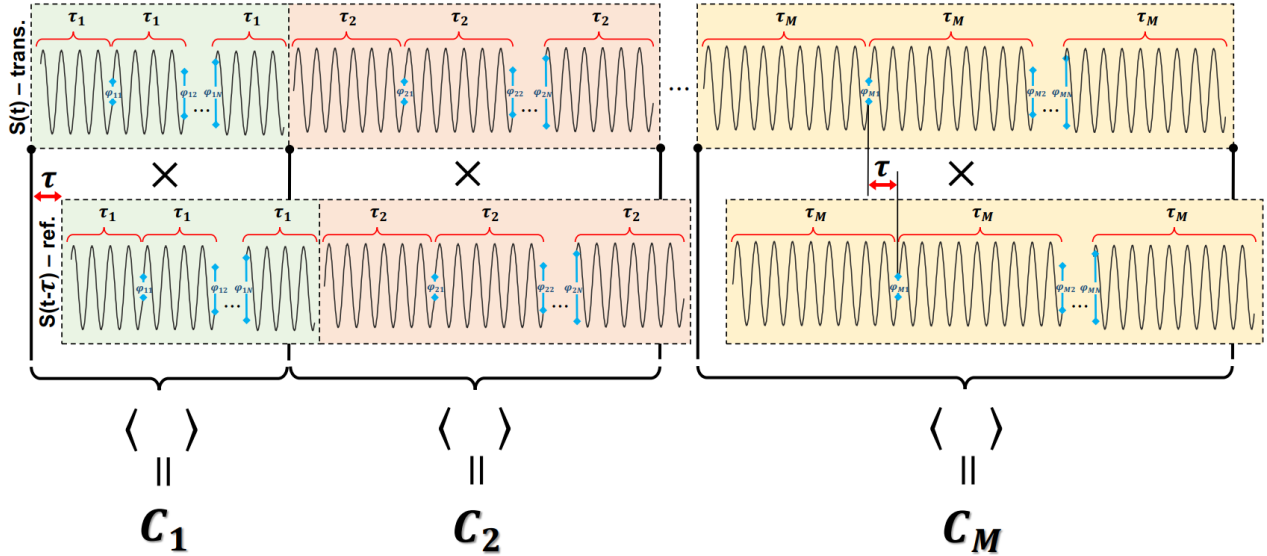

**Figure 2.** Illustration of the proposed detection method. The transmitted signal consists of a CW wave with a phase that is switched randomly every $\tau_m$ seconds, corresponding to opening a coherent window of length $c\tau_m$. If a target exists within the coherent window, the reflected signal will be delayed by time $\tau$ and can be divided into two parts – the first is of duration $(\tau_m - \tau)$, which is correlated with the still transmitting signal (i.e. the same phase), the other part is of duration $\tau$ and is uncorrelated with the transmitting signal. By switching the phase $N$ times and averaging the product of the reflected and transmitting signal over a window of length $N\tau_m$, the cross correlation $C_m$, which is the cross correlation for coherence time interval $\tau_m$, is measured. The cross correlation averages to 0 if the target is not within the coherence length. For a target within the coherent window, however, the average increases as the target becomes closer. By increasing the coherent window and repeating the process, the cross correlation as a function of coherence length can be obtained (see Fig.3).

The signal in Eq.1 can be thought of as a random phase pulse code, which is distinct from the well-known 'pseudo random noise' [21],[27], by effectively having 100% duty cycle (CW), as well as possessing completely random phases and varying PRF (pulse repetition frequency). Since in this implementation the coherence length corresponds to pulse width, the detection method could also be thought of as a type of 'pulse width modulation radar'.

**B.   *Single Stationary Target Detection***

The process of detection in the presence of a single stationary target (SST) is described in Fig. 2 and Fig. 3(a). For a given coherence time $\tau_m$ (which is related to coherence length $c\tau_m$), the expectation value of the cross correlation of the signal $C_m^{SST}$ with its echo, delayed by time $\tau$, attenuated by factor $A$ and immersed in white Gaussian noise $n(t)$, is given by:

$$E[C_m^{SST}] = E\left[\frac{A}{N\tau_m} \int_{T_m}^{T_m+N\tau_m} S(t)(S(t-\tau) + n(t))dt\right]. \tag{2}$$

Considering the case $\tau_m > \tau$ (i.e. the target is within the coherent window), the signal in Eq.1 is substituted into Eq.2 (see Supplementary Note 1). Noting that the signal is uncorrelated with the noise, the expression can be broken down into two parts, further simplified by trigonometric identities and the fact $\omega\tau_m \gg 1$ (see Fig.2 for an illustration):

$$E[C_m^{SST}] \stackrel{\omega\tau_m \gg 1}{\approx} \frac{A}{2N\tau_m} \sum_{n=0}^{N-1} E\left[\int_{n\tau_m}^{n\tau_m+\tau} Cos(\varphi_{n,m} - \varphi_{n-1,m} + \omega\tau)dt + \int_{n\tau_m+\tau}^{(n+1)\tau_m} Cos(\omega\tau)dt\right]. \tag{3}$$

The first integral is over the uncorrelated window, where $\varphi_{n,m} \neq \varphi_{n-1,m}$, and so the expectation value will vanish. The second term is over the correlated window and will contribute to the final result. For $\tau_m < \tau$, i.e. the target is not within the current coherent window, there will be no correlated part and the expectation value will vanish completely. It is convenient to rewrite the final result in terms of the spatial coherence length ($l_m = c\tau_m, l = c\tau$), where c is the speed of light, as well as defining $\tilde{C}_m^{SST} = l_m C_m^{SST}$ in order to linearize the dependence of the cross correlation on the coherence length:

$$E[\tilde{C}_m^{SST}(l_m)] = \begin{cases} \frac{A}{2}(l_m - l)Cos(kl) & \text{if } l_m > l \\ 0 & \text{otherwise} \end{cases}. \tag{4}$$

The standard deviation of the cross correlation around the mean can be calculated in a similar manner using the considerations leading up to Eq.3. In the absence of noise the variance is:

$$Var[C_m^{SST}] = Var\left[\frac{A}{2N\tau_m}\sum_{n=0}^{N-1}\left(\int_{n\tau_m}^{n\tau_m+\tau}\cos(\varphi_{n,m}-\varphi_{n-1,m}+\omega\tau)dt + \int_{n\tau_m+\tau}^{(n+1)\tau_m}\cos(\omega\tau)dt\right)\right] \quad (5)$$

This time the second integral does not contribute, as it is not a random variable. Since the different cosines in the sum are independent, the variance is simply the sum of variances:

$$Var[C_m^{SST}] = \sum_{n=0}^{N-1}Var\left[\frac{An\tau}{2N\tau_m}\cos(\varphi_{n,m}-\varphi_{n-1,m}+\omega\tau)\right] = \frac{A^2}{8N}\left(\frac{\tau}{\tau_m}\right)^2 \quad (6)$$

The above result considers $\tau_m > \tau$, otherwise there is no correlated part and the variance does not directly depend on the delay to the target (it still depends on the attenuation $A$, which increases with target distance). In addition, if the target were very far away, the delay would be so large that for the first transmitting 'pulses' there are no echoes to mix with, and so the variance in Eq.6 will be diminished. This, however, is not considered ahead in order to avoid cumbersome formulas. The noise term can be incorporated into the variance by introducing the signal to noise ratio at the receiving channel:

$$SNR = \frac{A^2 \frac{1}{T_M}\int_0^{T_M}|S(t)|^2 dt}{\sigma^2} = \frac{A^2}{2\sigma^2} \quad (7)$$

where $\sigma$ is the band limited power spectral density of the noise. The standard deviation can finally be written in full form using the above results:

$$\sigma_{C_m^{\text{SST}}} = \frac{A}{2}\begin{cases} \sqrt{\frac{1}{2N}\left(\frac{l}{l_m}\right)^2 + \frac{1}{SNR}} & \text{if } l_m > l \\ \sqrt{\frac{1}{2N} + \frac{1}{SNR}} & \text{otherwise} \end{cases}. \tag{8}$$

Eqs .4 and 8 allow to numerically estimate the cross correlation versus coherence length as shown in Fig.3, by assuming each point to be Gaussian distributed with a corresponding mean and standard deviation. The detection of targets is made by plotting the cross-correlation as a function of correlation length (or correlation time). The resulting graph is piece-wise linear and the location of the target corresponds to the breakpoint (see Fig.3(b and c) for the theory and for the experiment hereafter), which can be retrieved with the help of linear regression methods that were developed for the needs of stock market analysis [28]. Briefly, the approach here is to assume some point to be the breakpoint and then calculate the least squares fit to a linear function for the data on the right and separately on the left of that point. Repeating the process for all points and choosing the one that produced the least squares allows finding the breakpoint. More complex methods may be applied to further increase the accuracy of the breakpoint estimation as well as decrease the computational cost.

It's important to note that the choice of the carrier frequency is important in avoiding cross correlation zeroes when $\text{Cos}(kl) = 0$ in Eq.4, as can be seen on Fig.3 (b). This encourages an additional sweep over the carrier (frequency hopping for example) to insure detectability at any range. This additional sweep bandwidth is at most $\Delta f = \frac{1}{2\tau_0}$, which is between 5-7.5MHz for targets that are 15-10m away from the radar, as is the case in our experiment and future relevant applications. This additional bandwidth, which becomes even smaller for targets located further away, will be shown to be a relatively small addition. The impact of the carrier frequency choice on the cross correlation is shown on Fig 3(b) – while the location of the breakpoint does not change, the slope of the correlation function strongly depends on the carrier.

### C.   Multiple Stationary Targets Detection

For multiple stationary targets (MST) the expectation value of the cross correlation becomes a sum. Using similar arguments as for the single target, the cross correlation can be reduced to a sum of single target terms in the following form:

$$\mathrm{E}\big[\tilde{C}_m^{MST}(l_m)\big] = l_m \sum_{1}^{K} A_i \int_{T_m}^{T_{m+1}} \mathrm{E}[S(t-\tau_i)(S(t)+n(t))]dt = \begin{cases} \sum_{i\in D} \frac{A_i}{2}(l_m - l_i)\mathrm{Cos}(kl_i) \\ 0 \ \ if \ D = \emptyset \end{cases}, \quad (9)$$

where $A_i$ is the attenuation related to the distance and scattering cross section of the $i^{th}$ point-like target, $l_i = c\tau_i$ is twice the physical distance to target $i$, c the speed of light, $K$ the number of targets and $D = \{i: l_m > l_i\}$ is the set of all target indices that are within the coherence length of the signal. The standard deviation at each coherence length point $m$ will depend on the amount of targets within the coherence length. Since the echoes are additive, the standard deviation in the presence of $K$ targets is:

$$\sigma_{\tilde{C}_m}^{MST}(l_m, \{l_i\}_{i=1}^K, \{A_i\}_{i=1}^K) = \sqrt{\sum_{i=1}^{K} \left(\sigma_{\tilde{C}_m}^{SST,i}\left(\frac{l_i}{l_m}, A_i\right)\right)^2}. \quad (10)$$

Eq. 9 reveals that for multiple targets along the line of sight, several breakpoints on the cross-correlation graph are expected to appear. The location of the breakpoints on the plot identifies the physical position of the reflecting targets (see Fig. 3(d) for the theoretical plot and for the experimental result hereafter). The distance between the targets is half the distance between the breakpoints due to the monostatic operation of the radar. By performing a finer scan of the range and choosing an appropriate frequency that avoids cross correlation nulls for both targets, it is theoretically possible to distinguish uncoupled point targets located arbitrarily close to each other. This separation is only limited by the standard deviation of Eq.10, which can be made arbitrarily small by increasing the number of phase switches $N$ and the SNR. As will be shown in the following section, the required bandwidth depends entirely on the initial scanned range (the closest point scanned on the line of sight), meaning that the range resolution of the proposed system (the ability to distinguish close targets) does not depend on bandwidth.

D. *Sweep-Time and Bandwidth Tradeoff*

For $N$ pulses, $M$ coherence sweep points that begin at $\tau_0$ and scan a range $\Delta\tau$ of coherence time, the total scan time is

$$T_{tot} = N \sum_{m=0}^{M-1} \tau_m = \frac{(2\tau_0 + \Delta\tau)}{2} NM. \qquad (11)$$

Eq.11 shows a trade-off between the range accuracy and total sweep time, where a good precision requires high M and N, which prolongs the scan time. The maximal transmitted signal bandwidth (defined as the spectral distance between zeroes in the 'sinc' function, which is the Fourier transform of the rectangular envelopes, defined in Eq. 1) depends entirely on the starting coherence length time, $BW_{max} = \frac{2}{\tau_0}$, allowing to rewrite Eq. 11 in the following form:

$$BW_{max} = \frac{2}{\frac{T_{tot}}{NM} - \frac{\Delta\tau}{2}}. \qquad (12)$$

Eq.12 shows clearly that the proposed partially coherent radar is trading maximal transmitted bandwidth for sweep time, showing an inverse dependence.

### E. Effects of Moving Targets

In order to account for the effects of moving targets, the delay $\tau$ in Eq.2 should be replaced with a function of time. Assuming a single moving target (SMT) with speed *v* along the line of sight, the delay between the transmitted and received signals is now $\tau(t) = \tau + 2\frac{v}{c}t$, where the factor of 2 is due to the back and forth travel time of the wave, accounting for the classical Doppler effect in monostatic radars. Since the coherence time $\tau_m$ is short (typically less than a microsecond) it can be assumed, as normally done in radar analysis, that the target remains stationary during this period and that the Doppler effect results merely in the phase accumulation between adjacent 'pulses'. Further simplification can (but does not have to) be made by assuming that the target does not move much during the sweep-time-per-point $N\tau_m$, and that the target only changes its range when the coherence length is switched at the next iteration *m*. The last simplification allows obtaining a compact solution, which otherwise will be cumbersome. This approximation can be justified by considering low target velocities of around 200km/h, and a sweep-time-per-point of under a millisecond (corresponding to N<10,000). In such a case the target will have moved by about 6cm, which is around the accuracy of many radars including the experimental system described ahead. Finally, an assumption about the scanning mechanism needs to be made. For simplicity, consider a sequential scanning algorithm, which starts from the shortest coherence length and monotonically increases towards the final length scanned along the line of sight. In such a scenario, the target location is moving monotonically along the line of sight, presenting a different delay at each coherence length. The above discussion is applied to Eq.3 by replacing $\tau \to \tau + 2n\tau_m \frac{v}{c}$ inside the integral and considering only the correlated part. When the target is within the current coherence length, I.e. $\tau_m > \tau + 2\underbrace{\frac{(M-1)(\tau_m-\tau_0)}{\Delta\tau}}_{m} N\frac{v}{c}\tau_m$ the equation reads:

$$E[C_m^{\text{SMT}}] \overset{\omega\tau_m \gg 1}{\approx} \frac{A}{2N\tau_m} \sum_{n=0}^{N-1} E\left[\int_{n\tau_m+\tau}^{(n+1)\tau_m} \cos\left(\omega\left(\tau + 2\frac{v}{c}n\tau_m\right)\right) dt\right]. \quad (13)$$

As before, if the target is not inside the coherent window, the cross correlation is expected to vanish. The solution to Eq. 13 could now be written in terms of range rather than delay and multiplied by the coherence length as done in Eqs. 4 and 9:

$$E[C_m^{\text{SMT}}] = \begin{cases} \frac{A}{2}(l_m - l)\cos\left(k(l + \frac{(N-1)v}{c}l_m)\right)\frac{\sin\left(N\frac{kl_mv}{c}\right)}{N\sin\left(\frac{kl_mv}{c}\right)} & \text{if } l_m > l + 2mN\frac{v}{c}l_m \\ 0 & o.w \end{cases} \quad (14)$$

For low velocities the term $\frac{\sin\left(N\frac{kl_mv}{c}\right)}{N\sin\left(\frac{kl_mv}{c}\right)}$ in Eq.14 tends to unity, leaving a solution that is similar in form to Eq.4. The difference is found in the appearance of an oscillatory term as a function of coherence length $l_m$, which is due to phase accumulation between constant coherence iterations $n$, as well as an updated condition that considers the movement of the target between varying coherence length iterations $m$. Fig. 3(d) depicts the cross correlation as a function of coherence length for a moving target at different speeds, starting at a distance of 30m. For low velocity (36 km/h) the linear increase following the breakpoint deviates slightly from the stationary case.

When the velocity increases further (70 and 200 km/h) the breakpoint can be seen to recede from the stationary solution, due to the movement of the target, as described by the updated condition in Eq.14.

Eq.14 proves that the proposed detection method would only need a small adjustment in order to cope with slowly moving targets, by fitting the cross correlation as a function of coherence length to an oscillatory, rather than linear, function. Under the above approximations, the movement of the target has no effect on the standard deviation around the expected value, as can be deduced by adding a phase factor to the cosine in Eq.6. The same arguments that were used to derive Eq.14 could be applied together with the considerations leading up to Eq.9 and 10 in order to derive the cross correlation of multiple moving targets. Finally, it is important to note that the phase changes between adjacent 'pulses' of constant coherence time could be used in order to estimate the velocity of the target in the same manner as performed by standard pulsed radars [8], allowing for a quick estimation of the velocity before the end of the full coherence sweep. This velocity may be used in order to better fit the target location using Eq.14 and extrapolation formulas in future applications.

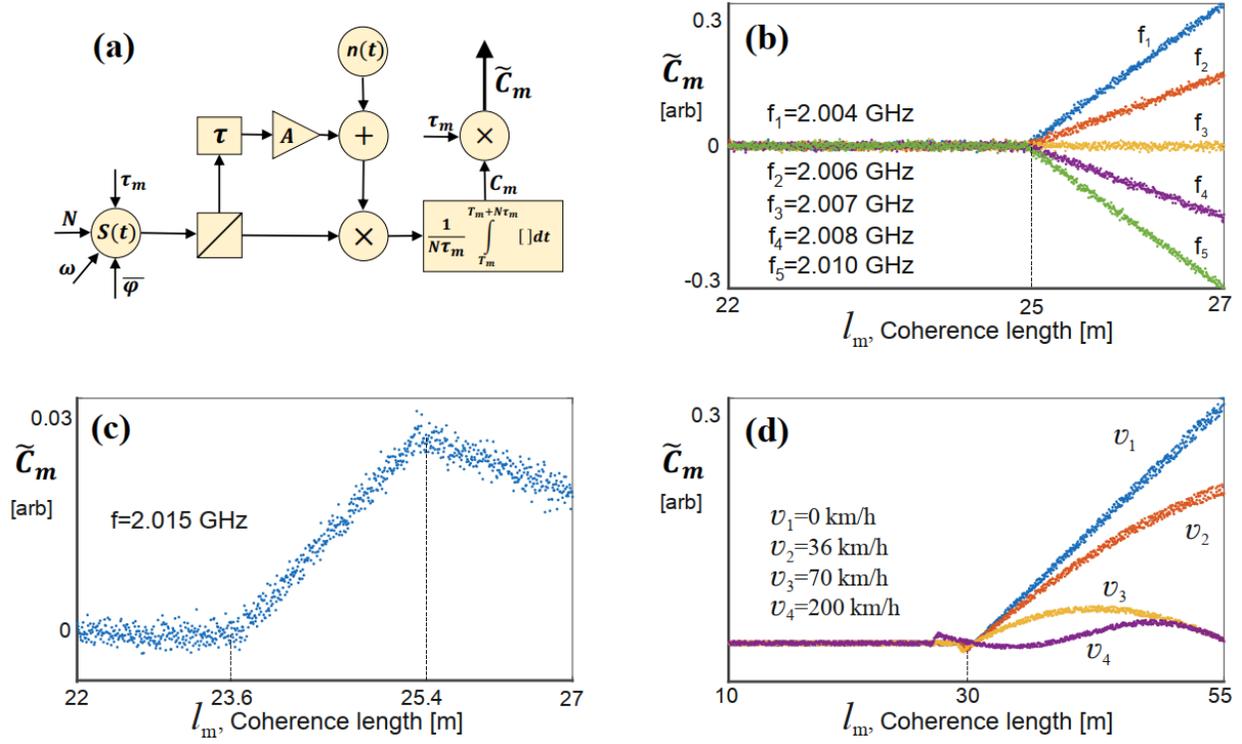

**Figure 3**. Simulation results and schematics (a) Schematic representation of the partially coherent radar operation. A continuous wave signal is generated with *N* phase jumps that are randomly produced to provide a constant (controllable) coherence time $\tau_m$ (time between phase switching events). Each pulse has a random phase ($\varphi_n$, or vector $\bar{\varphi}$), which is kept constant for the pulse duration. The signal reflects from a target that is situated at a distance related to the delay $\tau$, attenuated by a factor *A* and received along with additive white noise. The output of the receiver is mixed with the still transmitting signal and averaged over the duration of the transmission time $N\tau_m$ (the averaging starts at the same time as the signal begins transmitting). The result of the integration is multiplied by the coherence time $\tau_m$. The result is termed the cross correlation and denoted as $\tilde{C}_m$. The process is repeated for M coherence points (lengths of constant coherence). M and N define the performances of the system (range resolution and range accuracy). Monte Carlo simulations in high SNR (30dB) scenarios: (b) Cross correlation as a function of coherence length for a single target located 25 meters away, drawn for different carrier frequencies using Eq.4. (c) cross correlation for two targets located at coherence lengths of 23.6 and 25.4 meters corresponding to Eq.9. (d) Cross correlation of a single target moving at different velocities along the line of sight, corresponding to Eq.14.

## F.  Experimental Results

In order to demonstrate the performance of the system, a pair of scatterers were placed one in front of the other on the line of sight inside an anechoic chamber. Square plates were chosen in order to avoid ambiguity in measuring the distance between the targets, which was 32cm with the first object placed 2m away from the transmitting antenna. The transmitting antenna was connected with a long cable to the radar system, adding additional delay to the target in order to reduce the required bandwidth. Note, that a cable adds a deterministic effective distance to the target, and hence it effectively increases the duration of probing 'pulses', reducing the bandwidth as shown in Eq.12. The coherence length was swept electronically from 22 to 27 meters with *M*=500 coherence length points and *N*=5000 phase jumps per point, recording the cross correlation in the process. The carrier frequency was chosen to avoid zeros in the cross correlation (see discussion at the end of section A "Single Stationary Target" in "theory and implementation"). Using Eq.11 the sweep time is 204ms under optimal conditions, which can be reduced by performing less phase jumps and taking fewer sweep points, as well as implementing an advanced search algorithm (for example a binary search instead of the brute force sweep). Fig. 4(a) shows the photograph of the practical implementation of the partially coherent oscillator. This implementation was chosen over a simple phase shifter to allow for a faster switching time of the phase. Panel b demonstrates the actual transmitted bandwidth of the signal, which was used to obtain the experimental results depicted on Fig. 5(a).

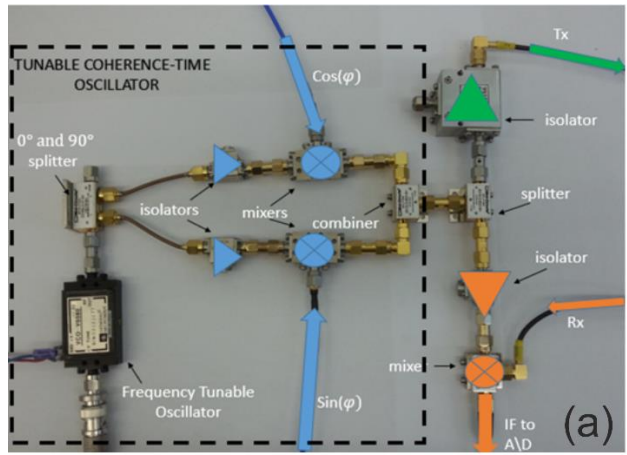

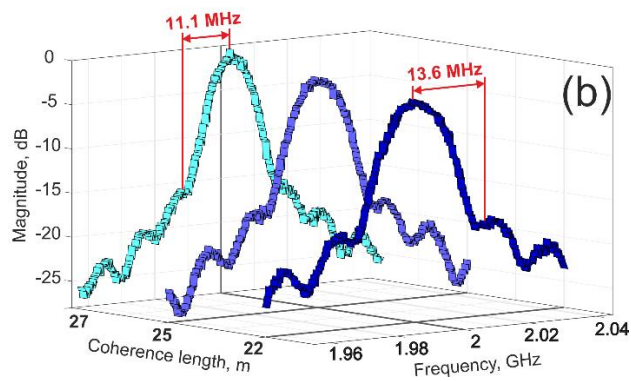

**Figure 4.** Experimental implementation and transmitted bandwidth measurements. (a) Photograph of the partially coherent oscillator implementation. A carrier is split into two quadratures, where one arm is phase delayed by 90 degrees. Each arm is multiplied by an appropriate sine or cosine term of the desired random phase output at time *t*, and the two arms are combined together to reveal a single carrier with the desired phase jumps, implementing and IQ vector modulator. (b) Measured half-bandwidth of the radar as it sweeps from coherence length of 22 to 27 meters, corresponding to transmitted bandwidths of between 27.2 – 22MHz. The peaks of the 'sinc' function are lowered with increasing bandwidth, conserving the transmitted power throughout the sweep. These signals were used to exploit the scenarios demonstrated in Fig. 5(a).

Fig. 5 shows the results for scenarios where either a single target, no target or both targets are present. When no target is present, a slight downward slope can be observed due to a small DC bias present in the experiment, which is due to imperfect isolation between the transmitting and receiving antennas, as well as reflections from various components. In Fig. 5(b and c) it is observed that for coherence lengths shorter than the distance to the target, the line is slowly sloping downward due to the same DC bias. As soon as the coherence length of the source achieves the back and forth distance to the target, a breakpoint occurs and the cross correlation between the transmitted and received signals starts to rise linearly, as predicted by Eq. 4 and shown on the theoretical Fig.3 (b). The different slopes in Fig. 5(b and c) are the result of target location and illumination frequency, determining the sign of $\text{Cos}(kl)$ for each target in Eq. 4. In Fig. 5(d), one can see the response when both targets are present, with slopes corresponding to the previous figures and in accordance with Eq.10 and the theoretical Fig.3(c). The distance between the resulting breakpoints is interpreted as twice the distance between the targets. By repeating the sweep numerous times and obtaining the location of the breakpoints each time, it is possible to estimate the probability densities of target range, shown as Gaussian insets in Fig. 4. The standard deviation of the probability density is the accuracy of the range, which is obtained to be about 10 cm. The distance between the targets from Fig. 5(d) is calculated as 35 cm, which is close to the actual physical distance (32 cm). Remarkably, when both targets are present, a shift of the first target's range can be observed from Fig. 5(b and d), which is the result of multiple reflections between the closely placed objects, which isn't considered in the simple point target model leading up to Eq.4 and 9. Fig. 4(b) shows the measured spectrum of the signals at chosen coherence lengths, showing the maximum transmitted bandwidth to be 27.2MHz. While this bandwidth is enough to resolve two targets 32 cm apart using the proposed method, the same bandwidth would only suffice to resolve targets separated by several meters using FMCW or pulsed radars, which is an improvement of more than an order of magnitude over standard systems. This improvement can be increased if the targets are located further away, as the highest bandwidth depends entirely on the starting point of the coherence length sweep. The starting point of the scan can be made arbitrarily far away artificially, by using delay lines (here coaxial cables were used as discussed before), finally untying range resolution from bandwidth limitations.

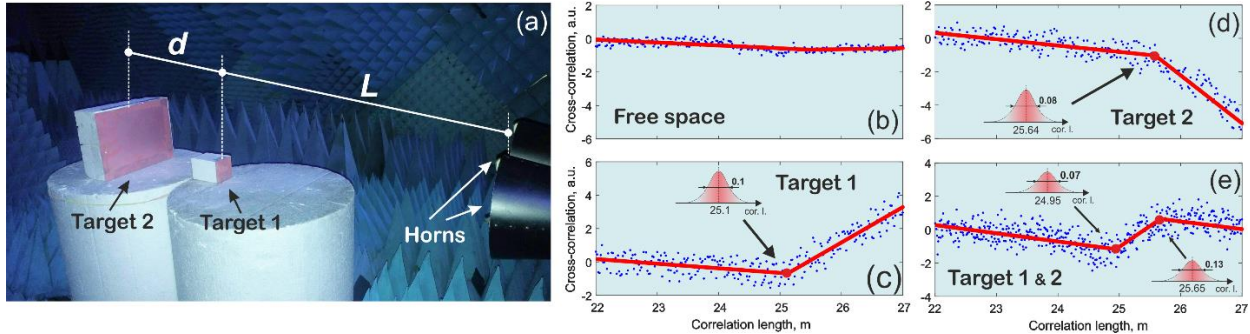

**Figure 5.** Detection and separation of two objects with narrowband signals – more than an order of magnitude below standard limitations. (a) Photograph of the experimental layout where two square plates were placed at a separation of 32cm away from each other. (b)-(e) Plots of cross-correlation as a function of the coherence length. Experimental data – blue points, piecewise linear fit – red solid lines. Insets – uncertainty in target locations. (b) Free space scan - empty room. (c),(d) single targets without the presence of the other, in accordance with the theoretical plot in Fig.3(b). (e) Both targets present at the same locations as before. The measured distance between the targets is 35cm, and it is close to the actual physical value (32cm) . The correlation length in (b)-(d) includes the physical distance to targets (as seen in (a)) as well as added distance due to cables and delays in other electronic components.

G. Comparison between Partially Coherent Radar and existing approaches

The vast majority of modern radar signals can be divided into two main groups: continuous wave (CW), transmitting continuously, and pulsed signals that transmit for relatively short periods. While both techniques improved significantly in the last few decades, none of them meet all the demands in the field of target detection and radar imaging owing to inherent limitations and trade-offs.

The majority of CW radar implementations are frequency modulated continuous wave (FMCW) radars, which are widely used in many applications today. The detection technique is based on mixing between a transmitted chirp signal and the received echo, which allows estimating the distance and velocity of a target [29] . FMCW radars are implemented all across the frequency spectrum [29],[30]. Many chirp modulation schemes are available (with increasing and then decreasing frequency, saw tooth modulation and others), however additional smart signal processing algorithms are needed to prevent severe problems of ghost target appearance [31]. FMCW schemes are commonly used in short range application in order to avoid the "blindness" problem that pulsed radars usually experience. Furthermore, there is an intrinsic relation between range resolution and bandwidth, demanding extremely expensive G (IEEE) band or even larger mm wave frequencies in order to separate several distant targets that are close to each other [32],[33].

Pulsed and pulsed Doppler radars are widely used for long distance detection, air traffic control being one of the main applications. These systems radiate short pulses and switch off the transmitter while waiting for echoes. This results in implementations that have an inherent blindness for short range targets. Compression techniques are used to cope with this problem with great success, but they still require large bandwidth and smarter signal processing at the receiver end [34].

Another type of ranging system that receives considerable interest is the noise radar [15], which does not fall into any of the beforehand mentioned classifications. The most basic type of noise radar cross correlates a random transmitted waveform with its received echo in order to determine the distance to the target. This approach has several advantages over conventional radars thanks to its random nature of electromagnetic radiation, which includes high immunity to noise and low probability of intercept which are relevant for military and urban applications. It does however demand high-precision controllable delay lines, which are expensive and hard to implement at mm waves as well as having high insertion loss [14]. Moreover, its range resolution still depends on the bandwidth, making it hard to implement an energy efficient and high range resolution noise radar [13],[16]. Table 1 presents a comparison between the commonly used radar implementations and the new Partially Coherent Radar.

| | **Inherent differences between common radar technologies and Partially Coherent Radar** | | | |
|---|---|---|---|---|
| | Pulsed Radar | FMCW Radar | Noise Radar | Partially Coherent Radar |
| Range Resolution dependence on bandwidth | $\frac{c}{2*BW}$ | $\sim \frac{c}{2*BW}$ Or even worse when using windowing techniques | $\sim \frac{c}{2*BW}$ | ✓ Free of bandwidth limitations |
| Short Range Target Detection | Requires large bandwidth for short range detection, and thus fast ADC (Analog-to-digital converter) [35] | ✓ Does not suffer from blind range due to its simultaneous transmit and receive scheme [29],[35] | Requires large bandwidth for short range detection, and thus very fast ADC, or low dynamic range [14],[35] | ✓ Does not require large bandwidth due to smart correlation detection algorithm |
| Long Range Target Detection | Demands compression techniques to be used for long range, and thus use larger bandwidth, making it more vulnerable to | Has a built in trade-off between range and resolution, which cannot be improved [29] | ✓ Has no restrictions on range | ✓ Has no restrictions on range |
| | manmade noise [36] | | | |
| Exploiting Doppler Effect for measurement of moving targets | ✓<br><br>Widely used to extract Doppler information. Fast targets don't severely affect range accuracy | ✓<br><br>Widely used to extract Doppler information. Fast targets don't severely affect range accuracy | Cannot be used for high speed targets [37] | ✓<br><br>Can be used to extract Doppler in the same manner as pulsed and FMCW radars. Fast targets (over 200km/h) can affect performance and require smarter algorithms |
| Other Limitations/Advantages | | Leakage from FMCW signal can impair the receiver especially at low received signal levels. Phase noise degrades performance [35] | Most applications require precise controlled delay lines, which have high insertion loss and are frequency depended [38] | ✓<br><br>Can be used even in sub 1GHz implementations |

*Table 1*. Comparison between widely used radar implementations and the Partially Coherent Radar

**Discussion**

. Statistical properties of electromagnetic radiation have been employed in order to remove the commonly accepted relation between the range resolution and the bandwidth of the transmitted signals. In particular, a new type of electromagnetic source with controllable coherence length was implemented and employed, demonstrating a ranging system possessing super resolution. Furthermore, this radar system is achieving a product of range resolution to bandwidth that is experimentally shown to be better by more than an order of magnitude compared with other radar technologies. In particular, it is shown that the range resolution can be virtually unrelated to the transmitted signal bandwidth, trading this quantity by the overall sweep time. The new system could be utilized to make bandwidth efficient, low power and physically compact systems for ranging purposes, integrated into existing beamforming and scanning systems. Therefore, autonomous cars, airborne radar systems, aerospace imaging along with other field of science and practical applications might exploit it to greatly improve their ability to detect targets and cope with the separation of close objects in densely populated frequency spectrum areas.

**Data availability**

The authors declare that [the/all other] data supporting the findings of this study are available within the paper [and its supplementary information files].

**Acknowledgments**

The research was supported in part by the ERC StG 'In Motion' (802279) and PAZY Foundation (Grant No. 01021248). The authors wish to thank Dr. Sergey Kosulnikov for his contribution.

**Authors Contribution**

R.K and V.K conducted the experiments and developed the theoretical framework, D.F analyzed the data and P.G developed the concept and supervised the work.

**Declarations**

The Authors declare no competing interests.